\documentclass{article}

\usepackage{microtype}
\usepackage{graphicx}
\usepackage{subfigure}
\usepackage{booktabs} 

\usepackage{hyperref}

\usepackage[accepted]{icml2023}

\usepackage{amsmath}
\usepackage{amssymb}
\usepackage{mathtools}
\usepackage{amsthm}
\usepackage{amsfonts}
\usepackage{bm}
\usepackage{graphicx,psfrag,epsf}
\usepackage{enumerate}
\usepackage{natbib}
\usepackage{float}
\usepackage{caption}

\usepackage[capitalize,noabbrev]{cleveref}

\def\*#1{\bm{#1}}

\theoremstyle{plain}

\theoremstyle{definition}

\theoremstyle{remark}

\usepackage[textsize=tiny]{todonotes}

\icmltitlerunning{Permutation-based Hypothesis Testing for Neural Networks}

\begin{document}

\twocolumn[
\icmltitle{Permutation-based Hypothesis Testing for Neural Networks}




\begin{icmlauthorlist}
\icmlauthor{Francesca Mandel}{penn}
\icmlauthor{Ian Barnett}{penn}
\end{icmlauthorlist}

\icmlaffiliation{penn}{Department of Biostatistics, Epidemiology, and Informatics, Perelman School of Medicine, University of Pennsylvania, Philadelphia, USA}

\icmlcorrespondingauthor{Francesca Mandel}{fmandel@pennmedicine.upenn.edu}

\icmlkeywords{hypothesis testing, inference, neural networks, machine learning}

\vskip 0.3in
]



\printAffiliationsAndNotice{}  

\begin{abstract}
Neural networks are powerful predictive models, but they provide little insight into the nature of relationships between predictors and outcomes. Although numerous methods have been proposed to quantify the relative contributions of input features, statistical inference and hypothesis testing of feature associations remain largely unexplored. We propose a permutation-based approach to testing that uses the partial derivatives of the network output with respect to specific inputs to assess both the significance of input features and whether significant features are linearly associated with the network output. These tests, which can be flexibly applied to a variety of network architectures, enhance the explanatory power of neural networks, and combined with powerful predictive capability, extend the applicability of these models.
\end{abstract}

\section{Introduction}
\label{section:intro}

While neural networks are well known for their predictive capability, compared to traditional regression approaches, they generally provide little explanatory insight into how they make their predictions. While the mathematics of each layer-to-layer transformation are relatively simple, how and why a network combines information from the inputs to predict the outputs becomes more difficult to understand as the network architecture grows in complexity. This issue of interpretability of neural networks has been addressed extensively in the literature \citep{gilpin2018explaining, zhang2021survey}. Despite the challenges, there are many settings in which is it desirable or necessary to interpret neural networks. In applications such as credit, employment, and criminal justice, understanding how predictions are made is extremely useful for evaluating whether the algorithms are fair and non-discriminatory \citep {bostrom2014ethics, hardt2016equality}. Recent laws mandating the ``right to explanation,'' a right to information about individual decisions made by algorithms, have accelerated the need for interpretability of complex models \citep{goodman2017european}. In many scientific research fields where data contain highly complex patterns, there is interest not just in accurate prediction but also in gleaning knowledge of the subject from the model fit. Machine learning methods are well-equipped to handle the size and complexity of genetic sequencing data, but improving network interpretability can further understanding of how novel variants contribute to susceptibility of diseases such as Parkinson's disease \citep{bryant2021identification}. Prognostic models for patients with severe brain injuries are critical to prescribing appropriate individualized treatment, and machine learning models that combine a variety of sources of information have shown great potential for enhancing the medical decision-making process. However, these models require a level of transparency and interpretability to be implemented in practice \citep{farzaneh2021hierarchical}.

Despite the well-documented need for interpretability with neural networks, there is not a clear consensus on what interpretability in this context means \citep{doshi2017towards, lipton2018mythos}. A wide variety of methods have been proposed to address different elements of interpretability. Many can be categorized as feature importance methods. Early work in this area included connection weights introduced in \citet{garson1991interpreting} and a saliency measure described in \citet{ruck1990feature}. \citet{dimopoulos1995use} proposed using partial derivatives to measure the sensitivity of a network and thereby determine its generalizability. Newer work has extended some of these ideas to more general concepts of feature relevance and explainability. \citet{bach2015pixel} proposed layer-wise relevance propagation (LRP), a framework for determining the relevance of input features in the determination of the network output. On an observation-by-observation basis, the output is propagated backward through the network according to a set of propagation rules that incorporate information from the weights and can be tailored to the network architecture and structure of the data. \citet{ribeiro2016should} introduced local interpretable model-agnostic explanations (LIME), a technique for explaining the predictions of any classifier or regressor, including neural networks, by learning an interpretable model locally around the prediction. DeepLIFT, an algorithm for assigning contribution scores to network inputs based on a difference-from-reference approach, was presented in \citet{shrikumar2017learning}. \citet{lundberg2017unified} unified these concepts in their 2017 paper and introduced Shapley additive explanation (SHAP) values, which quantify the contribution of each input feature for a particular prediction. \citet{sundararajan2017axiomatic} proposed integrated gradients as a measure of feature relevance. \citet{zhang2021survey} provides a useful survey of these and other methods. Several of these methods use some form of network gradients, however, their focus is on constructing measures of feature importance or interpretability rather than conducting formal tests of statistical significance.

A second category of methods aims to design network architectures that enable interpretability. \citet{potts1999generalized} proposed a generalized additive neural network (GANN), which fits a separate neural network with a single hidden layer for each input variable and combines the individual outputs. Leveraging advances in deep learning from the past decades, \citet{agarwal2021neural} developed neural additive models (NAM), which replace the smooth functions in generalized additive models (GAM) with deep neural networks. \citet{wojtas2020feature} introduced a dual-net architecture for population-level feature importance that simultaneously finds an optimal feature set that maximizes model performance and ranks the importance of the features in the subset. A selector network learns the optimal feature subset and ranks feature importance while an operator network makes predictions based on the optimal subset.

Developments in a third category of methods, significance testing of network inputs, have been more limited. \citet{olden2002illuminating} designed a randomization test for network weights that can be combined with the connection weights metric introduced by \citet{garson1991interpreting} to test for statistical significance of input features. \citet{horel2020significance} developed a test for the significance of input features in a single-layer feed-forward neural network. They proposed a gradient-based test statistic that is a weighted average of squared partial derivatives and studied its large-sample asymptotic behavior. \citet{racine1997consistent} addressed the issue of significance testing for nonparametric regression and devised a test based on partial derivatives with estimation of the null distribution via resampling methods. However, the test was designed for kernel regression rather than neural networks. Each of these methods focuses on testing whether associations exist between network inputs and outputs. Since neural networks can flexibly model complex nonlinear associations, it is of interest to extend the significance testing framework and study the nature of associations between inputs and outputs.

Hypothesis testing for neural networks offers several advantages over other interpretability methods. Many feature importance methods only provide local explanations of network behavior. Focusing on explainability at the individual prediction level can potentially obscure important information at the global level. When assessing the associations between network inputs and outputs, it is more desirable to take a global approach and account for the overall behavior of the network. Additionally, a feature importance ranking is only interpretable relative to the other network inputs. On the other hand, hypothesis testing provides a clear and objective interpretation of the significance of the inputs in predicting the network output. Methods that modify the network architecture to increase explainability can be powerful tools. However, the network architecture must still be compatible with the structure of the data and the type of interpretation desired, potentially limiting their usability in some settings. In contrast, the significance testing framework can be flexibly applied to general architectures.

Here we propose two hypothesis testing frameworks for evaluating the association between network inputs and outputs. The first test determines whether an input is nonlinearly associated with an output, and the second test evaluates the statistical significance of any type of association between each input feature and an output. In both tests, we construct a gradient-based test statistic and use permutation methods to estimate the null distribution. We use simulation studies to demonstrate the performance of our test under various types of data and compare to competing methods in the literature. Additionally, we apply our proposed tests to evaluate feature associations in pediatric concussion data and to test genetic links to Parkinson's disease.

The rest of the paper is organized as follows. In Section~\ref{section:methods}, we introduce the hypotheses, test statistics, and testing procedures. Section~\ref{section:simulations} evaluates the performance of our test relative to competing methods in simulation studies. In Section~\ref{section:data}, we apply our tests in two settings: pediatric concussion data and genomic data from Parkinson's patients. We conclude with discussions in Section~\ref{section:discussion}.

\section{Methodology}
\label{section:methods}

For notational simplicity, we henceforth assume a one-layer feed-forward neural network, but the approach is general and can be easily extended to more complex architectures. Consider the $i$th of $n$ observations with univariate outcome $y_i$ and vector $\*x_i = (x_{i1},...,x_{ip})^T$ of predictors. Let $\*X$ be the $n$ by $p$ matrix of predictors and $\*y$ be the vector of length $n$ of outcomes. Suppose a neural network with $p$ input features, a single hidden layer with $k$ nodes and nonlinear, differentiable activation function $g_1$, a univariate outcome $\mu_i$, and final layer activation function $g_0$ has been trained on $(\*x_i,y_i), i=1,...,n$. Of interest is the association between an input feature $\*X_j=(x_{1j},...,x_{nj})^T$ and outcome $\*y$. It is relevant to consider the partial derivative of the network output with respect to $\*X_j$. For a single-hidden-layer network with a univariate output, the partial derivative is

\begin{equation} \label{PaD}
\begin{split}
\frac{\partial \mu_i}{\partial x_{ij}} = & g'_0\left\{\*\omega^{(0)}\*\alpha_i^{(1)}+\delta^{(0)}\right\} \\
& \cdot \*\omega^{(0)} \left[ g'_1 \left\{ \*\omega^{(1)} \*x_i + \*\delta^{(1)} \right\} \odot \*\omega_{j \cdot}^{(1)} \right],
\end{split}
\end{equation}
where  $\*\omega^{(0)}$ are the final layer weights,  $\*\omega^{(1)}$ are the hidden layer weights, $\delta^{(0)}$ is the final layer bias, $\*\delta^{(1)}$ are the hidden layer biases, $ \*\omega_{j \cdot}^{(1)}$ is the vector $( \omega_{j1}^{(1)} \omega_{j2}^{(1)}...\omega_{jk}^{(1)})^T$, and $\odot$ is the Hadamard product. It is natural to assume that if the partial derivative in \eqref{PaD} is equal to 0 for all $\*x_i \in \*X$, then $\*X_j$ is not associated with $\*y$. Similarly, if the partial derivative is equal to a constant $c$ for all $\*x_i \in \*X$, then $\*X_j$ is linearly associated with $\*y$. This provides our motivation for basing our test statistic on this partial derivative. However, the partial derivative varies over the domain of $\*X$, so we must account for the wide range of values the test statistic can take. Furthermore, the asymptotic distribution of the partial derivative function is not easily derived. Instead, we rely on resampling techniques to estimate the null distribution of the test statistic. We outline the procedures for two tests: a test for nonlinear association between $\*X_j$ and $\*y$ and a test for general association between $\*X_j$ and $\*y$.

\subsection{Test for nonlinear association}
Suppose a network has been trained as described above. Of interest is whether a nonlinear association exists between input feature $\*X_j$ and outcome $\*y$. We can state the null hypothesis that  $\*X_j$ is linearly associated with $\*y$ in terms of the partial derivatives:
$$ H_0: \frac{\partial \mu_i}{\partial x_{ij}} = c \quad \forall \*x_i \in \*X $$
$$ H_A: \frac{\partial \mu_i}{\partial x_{ij}} \neq c \quad \text{for some}\; \*x_i \in \*X $$
for some constant $c$. We propose the following testing procedure. We first calculate the partial derivative in \eqref{PaD} for every observation in the data. Under $H_0$, the partial derivatives should be fairly constant across the domain of $\*X$. To evaluate whether the partial derivatives are sufficiently close to $c$, we calculate the residuals of the observed partial derivatives from their mean. We then fit a smooth function to the $n$ residuals and let the test statistic be the mean of the squared coefficients from the smooth function. To obtain a null distribution of the test statistic, we use networks trained on permutations of the observed data. We generate permutations of the data by permuting the model residuals of a GAM fit to $(\*X,\*y)$, where $\*X_j$ is restricted to a linear term in the model. In general, the test is robust to the specification of the smooth terms for the other $p-1$ variables. However, enough flexibility should be provided to reasonably capture the contribution of each input to the outcome variable. The permuted data consists of the observed predictors and a permuted outcome vector that is the sum of the fitted values from the estimated GAM and a permutation of the vector of model residuals. Permuting the data in this way forces the association between $\*X_j$ and $\*y$ to be linear while preserving any potential nonlinearity between the outcome and the other $p-1$ predictors. We then train a network on the permuted data and calculate the partial derivatives at every observation. We again fit a smooth function to the residuals of the partial derivatives and calculate the corresponding test statistic to be the mean of the squared coefficients. Under the null, the residuals of the partial derivatives will be randomly scattered around 0, so it should be the case that the smooth function is approximately 0 and therefore the test statistic is close to 0. Under the alternative, there will be a systematic pattern in the residuals, so the smooth function will be nonzero and the test statistic will be larger than 0. The p-value is then the proportion of the test statistics calculated under the null that are larger than the observed test statistic. For additional detail, see Appendix \ref{section:appendix}.

The test relies on the assumption that the neural networks are well-fitted to the data. Poorly trained networks may not accurately capture the true associations between predictors and the output, impacting the performance of the test. The test is fairly robust to the degree of smoothing; the estimated smooth functions should capture important patterns in the residuals without overfitting. Additionally, there are some implicit assumptions that arise from fitting a GAM in the permutation step of the test. First, GAMs are limited to modeling smooth effects, so any nonsmooth associations between predictors and the outcome may hurt the model fit and therefore affect the values of the permuted outcome vector. In practice, this may not have a large effect on test performance if the smooth estimate of the true nonsmooth association is reasonable. Second, unless explicitly specified in the model, GAMs cannot capture interactions like a neural network can. If there is knowledge or evidence of interaction effects, these can be included in the GAM used to permute the data. However, these should be restricted to predictors that are not being tested so the interpretation of the predictor of interest is not affected.

\subsection{Test for association}
Since the null hypothesis of the test for nonlinearity includes the possibility of no association between the input feature and the output of the network, it is of interest to test whether any type of association exists between $\*X_j$ and $\*y$. Specifically, we wish to test the following hypotheses: 
$$ H_0: \*X_j \text{ is not associated with } \*y $$
$$ H_A: \*X_j \text{ is associated with } \*y. $$
Alternatively, we can state the hypotheses in terms of the partial derivatives:
$$ H_0: \frac{\partial \mu_i}{\partial x_{ij}} = 0 \quad \forall \*x_i \in \*X $$
$$ H_A: \frac{\partial \mu_i}{\partial x_{ij}} \neq 0 \quad \text{for some}\; \*x_i \in \*X. $$

A resampling procedure similar to the nonlinearity test is used to test for an association. For a neural network trained on $(\*X,\*y)$, we calculate the partial derivative function in \eqref{PaD} for every observation in the data and let the observed test statistic be the mean of the squared partial derivatives. Under $H_0$, the partial derivatives should be approximately 0 across the domain of $\*X$. To obtain a null distribution of $T$, we use network fits based on permutations of the original data. We permute the vector of observed values of $\*X_j$ such that all columns of the new predictor matrix are identical to the original predictor matrix $\*X$ except the $j$th. By permuting the values of $\*X_j$, any potential association between the $j$th input and the output is erased, reflecting $H_0$. We then train a network on the permuted data, calculate the partial derivatives at every observation, and compute the test statistic to be the mean of the squared partial derivatives. We expect the partial derivatives to be close to 0 under the null, so the test statistic will be close to 0. Under the alternative, the partial derivatives should be nonzero, so the test statistic will be larger than 0. Then, the p-value is the proportion of the test statistics calculated under the null that are larger than the observed test statistic. We list the steps of the test in detail in Appendix \ref{section:appendix}.

It is important to note that the joint distribution of the predictors is broken by permuting $\*X_j$. Thus, a key drawback of this approach is the implicit assumption of independence among the predictors. At a minimum, the test assumes that the predictor of interest $\*X_j$ is not correlated with the other predictors, though the other $p-1$ predictors can be correlated with one another in an arbitrary pattern. The sensitivity of test performance to correlated predictors is explored empirically in Section~\ref{section:simulations}. Additionally, as with the test for nonlinearity, the neural network must be well-fitted to the data. If the network has not been trained well, the partial derivatives may not represent the true nature of the relationship between the predictors and outputs, and consequently the test may not perform well.

\subsection{Suggested usage of tests}
To fully characterize the association between $\*X_j$ and $\*y$, both the nonlinearity and association tests may be needed. If the test for nonlinearity is implemented first and there is evidence of nonlinearity, then no further testing is needed. However, if the test suggests there is a linear relationship, then the association test must be used to determine whether an association exists at all. Two unassociated variables can be said to follow the linear relationship $\*y \propto c \*X$ where $c=0$. Therefore, the nonlinearity test cannot be used alone to determine a nonzero linear association. Alternatively, if the association test is implemented first and the test suggests there is no association, then no further testing is required. If the test finds evidence of an association, the test for nonlinearity can then be used to determine the nature of that association. To implement the tests in practice, a network can be fit on the original observed data and used for both tests. Then the permutation of the data and retraining of the network can be run separately for each test.

\section{Simulation Studies}
\label{section:simulations}
We evaluate the performance of our proposed tests through several simulation studies. Where applicable, we include comparisons to competing methods.

\subsection{Power and Type-I error of nonlinearity test}
We estimate the power and Type-I error of the proposed test for nonlinearity through simulation. Let $i$ denote the observation. For $i = 1,...,500$, we generate five continuous independent variables $\*x_i = (x_{i1},x_{i2},x_{i3},x_{i4},x_{i5})^T$ from a standard normal distribution. A univariate continuous outcome is generated by the model $y_i = -\beta x_{i1} + \beta x_{i2}^2 - \beta x_{i3}^3 + \beta \sin(2x_{i4}) - \beta |x_{i5}| + \epsilon_i$, where $\beta = 0.2$ and $\epsilon_i \sim N(0,0.2)$, and a test of nonlinearity with 500 permutations is conducted for each of the five predictors. We fit a neural network with one hidden layer with 40 nodes and a sigmoid activation function. The network is trained for 150 epochs using stochastic gradient descent, $L_2$ regularization, and a decreasing learning rate at every epoch. The number of hidden nodes, the initial learning rate, and the regularization parameter are chosen to minimize loss on a validation set. Power and Type-I error are estimated across 300 simulations.

Power and Type-I error of the nonlinearity test are presented in Table \ref{tab:nonlinearity_test_simulation}. The estimated Type-I error rate is 0.05. The test has high power to detect a variety of nonlinear effects. The nonlinearity of the quadratic term ($X_2$) is easily detected by the test, with an estimated power of 1. Although the association between the cubic term ($X_3$) and the outcome could be reasonably approximated by a linear function, the test maintains high power in detecting the true nonlinearity with power equal to 1. The power of the test is slightly lower for the trigonometric ($X_4$) term, due to the complexity of modeling a periodic function. The test has high power even in the nonsmooth setting, with an estimated power of 0.94 for $X_5$. Overall, the test for nonlinearity performs well under a variety of alternatives, even when the association can be well approximated by a linear function. 

\begin{table}[t]
\caption{Power and Type-I error of the neural network permutation test for nonlinearity. The probability of rejecting $H_0$ is calculated at the 5\% level for five types of associations: linear, quadratic, cubic, trigonometric, and nonsmooth.}
\label{tab:nonlinearity_test_simulation}
\vskip 0.15in
\begin{center}
\begin{small}
\begin{sc}
\begin{tabular}{llr}
\toprule
Variable & Association & Pr(Reject $H_0$) \\ 
\midrule
$X_1$ & Linear & 0.05 \\
$X_2$ & Quadratic & 1.00\\
$X_3$ & Cubic & 1.00 \\
$X_4$ & Trigonometric & 0.86 \\
$X_5$ & Nonsmooth & 0.94 \\
\bottomrule
\end{tabular}
\end{sc}
\end{small}
\end{center}
\vskip -0.1in
\end{table}

\subsection{Power and Type-I error of association test}
We compare the performance of the proposed association test for neural networks to association tests from two traditional interpretable models: linear models and GAMs. The properties of standard $t$-tests from a linear regression model are well-established, however, these only hold under the assumption of linear associations between predictors and outcomes. GAMs can flexibly model many nonlinear trends, and testing for associations between predictors and outcomes is straightforward using the p-values for smooth terms outlined in \citet{wood2013p}. However, GAMs are limited to smooth effects \citep{hastie2017generalized}.

As discussed in the introduction, NAMs have been introduced as a way to combine the predictive capability of deep neural networks with the interpretability of GAMs \citep{agarwal2021neural}. The model's explainability is due to the ability to easily visualize how it computes a prediction. Since the impact of a feature on the predicted output does not rely on the other network inputs, each feature can be assessed individually by plotting its estimated shape function. While the architecture of NAMs makes them far easier to interpret than standard deep neural networks, explainability is still based on subjective interpretation of a graph. Therefore, we view NAMs not as a competing method, but as a model to which our proposed tests could be applied, and as such, we do not include them in our simulation studies.

To compare the performance of testing for association in neural networks, GAMs, and linear models, power and Type-I error under various settings are estimated through simulation. Three data generation mechanisms are considered: linear, smooth nonlinear, and nonsmooth nonlinear. Let $i$ denote the observation. For each setting, four continuous independent variables $\*x_i = (x_{i1},x_{i2},x_{i3},x_{i4})^T$ are generated from a standard normal distribution. In the linear setting, a univariate continuous outcome is generated from the linear model $y_i = \*x_i^T \*\beta + \epsilon_i$, where under the null, $\beta_j \sim N(0.3,0.01^2),\ j=1,2,3$ and $\beta_4=0$, and under the alternative, $\beta_j \sim (m,0.01^2),\  j=1,2,3,4$. The mean $m$ takes values $\{ 0.24,0.27,0.30,0.33,0.36 \}$. In the smooth nonlinear setting, a univariate continuous outcome is generated from the model $y_i = \beta_1 x_{i1}^3 + \beta_2 \cos(x_{i2}) + \beta_3 \tanh(x_{i3}) + \beta_4 \sin(3x_{i4}) + \epsilon_i$, where under the null, $\beta_j \sim N(0.3,0.01^2),\ j=1,2,3$ and $\beta_4=0$, and under the alternative, $\beta_j \sim (m,0.01^2), \ j=1,2,3,4$. The mean $m$ takes values $\{ 0.24,0.27,0.30,0.33,0.36 \}$. In the nonsmooth nonlinear setting, $\*x_i$ defines a vector $\*z_i = (z_{i1},z_{i2},z_{i3},z_{i4})^T$, where 
$$z_{i1}= \begin{cases}
x_{i1}x_{i2} & x_{i1}x_{i2}<0 \\
0 & x_{i1}x_{i2} \geq 0 
\end{cases} \; z_{i2}= \begin{cases}
x_{i2}x_{i3} & x_{i2}x_{i3}>0 \\
0 & x_{i2}x_{i3} \leq 0 
\end{cases}$$

$$z_{i3}= \begin{cases}
x_{i3}x_{i4} & x_{i3}x_{i4}<0 \\
0 & x_{i3}x_{i4} \geq 0 
\end{cases} \; z_{i4}= \begin{cases}
x_{i4}x_{i1} & x_{i4}x_{i1}>0 \\
0 & x_{i4}x_{i1} \leq 0.
\end{cases}$$
A univariate continuous outcome is generated from $y_i = \*z_i^T \*\beta + \epsilon_i$, where under the null, $\beta_j \sim N(0.18,0.01^2),\  j=1,2,3$ and $\beta_4=0$, and under the alternative, $\beta_j \sim (m,0.01^2),\  j=1,2,3,4$. The mean $m$ takes values $\{ 0.12,0.24,0.36,0.48,0.60 \}$. In all settings, $\epsilon_i \sim N(0,0.3)$. 500 observations are generated for each setting.

We perform a test of association between the predictor $\*X_4$ and the outcome $\*y$ in each setting. For the linear model (LM), a standard linear regression is fit on all four predictors, and the p-value from a $t$-test for $\beta_4=0$ is used. GAMs are fit with a separate smooth term with a 10-dimensional cubic regression spline basis for each predictor, and p-values for significance of the smooth term for $\*X_4$ are calculated as described in \citet{wood2013p}. We fit neural networks with one (NN-1) and two hidden layers (NN-2) and sigmoid activations. We train using stochastic gradient descent, $L_2$ regularization, and a decaying learning rate. The number of nodes, the initial learning rate, and the regularization parameter minimize loss on a validation set. The proposed test for association is conducted with 500 permutations. Power and Type-I error are estimated across 500 simulations.

\begin{figure}[ht]
\vskip 0.2in
\begin{center}
\centerline{\includegraphics[width=\columnwidth]{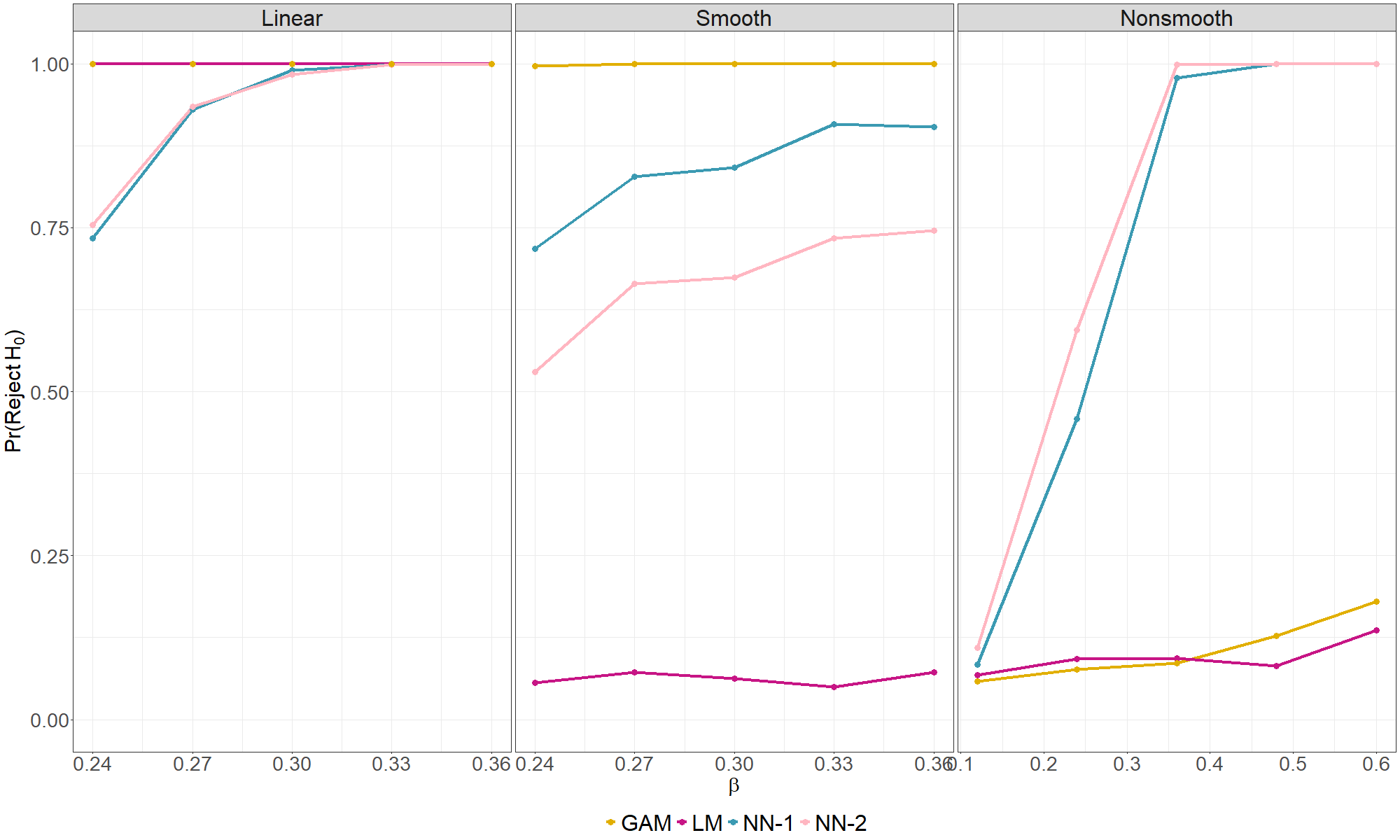}}
\caption{Power curves of tests for association using neural networks (NN-1 and NN-2), linear models (LM), and generalized additive models (GAM). Data is generated from three models (linear, smooth nonlinear, and nonsmooth nonlinear) and five signal levels. Each point is the rate of rejecting $H_0$ (based on $p \leq 0.05$ for NN-1, NN-2, and LM, $p \leq 0.035$ for GAM) from 500 simulations.}
\label{fig:assoc_tests_power}
\end{center}
\vskip -0.2in
\end{figure}

Table \ref{tab:assoc_tests_t1} contains the estimated Type-I error rates, and Figure \ref{fig:assoc_tests_power} shows the power curves for the three data generation models and four testing methods. Due to the conservative nature of the p-values for smooth terms in a GAM, the significance threshold was adjusted to 0.035 for these models to allow fair comparison among the methods. Type-I error of the permutation test is accurate across all settings, though slightly conservative in the smooth and nonsmooth settings. In the linear setting, all methods perform similarly, an expected result given that linear models, GAMs, and neural networks can all estimate linear effects. LM and GAM slightly outperform the permutation test under low signal, likely due to a loss of efficiency from fitting a neural network with a large parameter space for a simple linear association. In the smooth setting, power is significantly lower for LM since it is limited to modeling linear effects while power remains high for both GAM and NN. In the nonsmooth setting, our proposed test significantly outperforms the competing methods. This is expected as linear models and GAMs cannot adequately model nonsmooth nonlinear associations. In contrast, these complex relationships are easily learned by neural networks, and therefore the proposed permutation test can accurately detect the presence of associations between the predictors and outcome. At small signal levels, the added complexity of a second hidden layer in NN-2 results in a slight improvement in power compared to NN-1.

\begin{table}[t]
\caption{Type-I error of tests for association using neural networks (NN-1 and NN-2), linear models (LM), and generalized additive models (GAM).  Data is generated from three models (linear, smooth nonlinear, and nonsmooth nonlinear). Type-I error is the rate of rejecting $H_0$ (based on $p \leq 0.05$ for NN-1, NN-2, and LM, $p \leq 0.035$ for GAM) from 500 simulations.}
\label{tab:assoc_tests_t1}
\vskip 0.15in
\begin{center}
\begin{small}
\begin{sc}
\begin{tabular}{lrrr}
\toprule
 & Linear & Smooth & Nonsmooth \\ 
\midrule
NN-1 & 0.030 & 0.026 &  0.028\\
NN-2 & 0.056 & 0.024 &  0.026 \\
LM & 0.056 & 0.050 &  0.048 \\
GAM & 0.048 & 0.038 &  0.050\\
\bottomrule
\end{tabular}
\end{sc}
\end{small}
\end{center}
\vskip -0.1in
\end{table}

In real data settings, it is likely the predictors may be correlated. We assess the degree to which correlation among the network inputs impacts the Type-I error of the association test. We draw 500 observations of $\*x_i = (x_{i1},x_{i2},x_{i3},x_{i4},x_{i5},x_{i6},x_{i7},x_{i8})^T$ from a multivariate normal distribution with correlation $\*\Sigma$. We consider three settings for $\*\Sigma$: independence, low correlation, and high correlation. We choose $\*\Sigma$ to reflect real data structures by selecting the low and high correlation settings from the empirical correlation matrix of a subset of features from the Children's Hospital of Philadelphia pediatric concussion data, described in Section \ref{section:data}. The low correlation matrix is estimated from eight elements of the Sport Concussion Assessment Tool and has a mean magnitude of correlation of 0.13, and the high correlation matrix is estimated from eight elements of the Post-Concussion Symptom Inventory and has a mean magnitude of correlation of 0.60. For each of the 500 observations of $\*x_i$, we generate $y_i$ from the model $y_i = \beta x_{i2}^2 + \beta \cos(x_{i3}) + \beta \sin(2x_{i4}) + \beta x_{i5} + \beta x_{i6} + \beta x_{i7} + \beta x_{i8} + \epsilon_i$, where $\beta \sim N(0.1,0.01^2)$ and $\epsilon_i \sim N(0,0.1)$. We conduct an association test for $\*X_1$ with 500 permutations. A one-layer network with 30 nodes and sigmoid activation is trained for 150 epochs using stochastic gradient descent, $L_2$ regularization, and a decaying learning rate. The number of nodes, the initial learning rate, and the regularization parameter minimize loss on a validation set. Power and Type-I error are estimated across 300 simulations.

With independent predictors, the estimated Type-I error rate is 0.05. However, this rate increases to 0.09 under low correlation and to 0.32 under high correlation. The diminished performance of the test under correlated settings is a direct result of breaking the joint distribution of the predictors in the permutation step of the testing procedure. The increase in Type-I error is moderate under low correlation but very large under high correlation. These results suggest the proposed test is best implemented when there is a reasonable assumption of near-independence among the predictors.

\section{Data Applications}
\label{section:data}

We apply the permutation tests in two settings. Our first application uses pediatric concussion data from the Center for Injury Research and Prevention at the Children's Hospital of Philadelphia (CHOP) to assess associations between clinical and device-based diagnostic measures \citep{corwin2021visio}. Our second application uses genomic data from the Accelerating Medicines Partnership Parkinson's Disease (AMP PD) project (2019 v1 release) to test for associations of genes linked to Parkinson's with disease status.

\subsection{Testing associations among concussion diagnostics}
Teenage participants from a suburban school were enrolled in a large, prospective observational cohort study assessing various diagnostic measures of concussion. Concussed subjects sustained sport-related injuries; non-concussed subjects completed testing as part of an assessment for a scholastic sport season. The Sport Concussion Assessment Tool, $5^{th}$ Edition (SCAT-5) is a concussion assessment battery that measures symptom burden, memory, and concentration \citep{echemendia2017sport}. We consider three variables from SCAT-5: symptom score of 0-22 and symptom severity score of 0-132 from assessing 22 symptoms on a 7-point scale, and delayed word memory, where subjects repeat a list of five words after five minutes elapse. The Post-Concussion Symptom Inventory (PCSI) is a self-report questionnaire of symptoms rated on a 7-point scale \citep{sady2014psychometric}. We consider two individual elements of the PCSI (headache and dizziness) and two combined scores (emotional symptom score and total symptom score). Total symptom score is the sum of each individual symptom, including headache and dizziness. Pupillary light reflex (PLR) metrics can potentially assess visual dysfunction following concussion \citep{master2020utility}. We consider two PLR metrics: average pupil constriction velocity (ACV) and time for pupil redilation to 75\% of maximum diameter (T75).

We analyze a data set of 544 observations including cases and controls. We fit two separate one-layer networks with sigmoid activations and $L_2$ regularization. The first network (NN-SCAT) predicts SCAT-5 symptom score with four inputs: SCAT-5 symptom severity score, SCAT-5 delayed word memory, PCSI emotional score, and age. The second network (NN-PCSI) predicts PCSI total score with four inputs: PCSI headache, PCSI dizziness, PLR ACV, and PLR T75. We employ stochastic gradient descent with an initial learning rate of 0.005 for NN-SCAT and 0.01 for NN-PCSI; the learning rates decay by 1.5\% at each epoch. The number of nodes and the regularization parameter minimize validation loss. Both networks have 20 nodes and a regularization parameter of 0.03. The networks train for 175 epochs.

\begin{figure}[ht]
\vskip 0.2in
\begin{center}
\centerline{\includegraphics[width=\columnwidth]{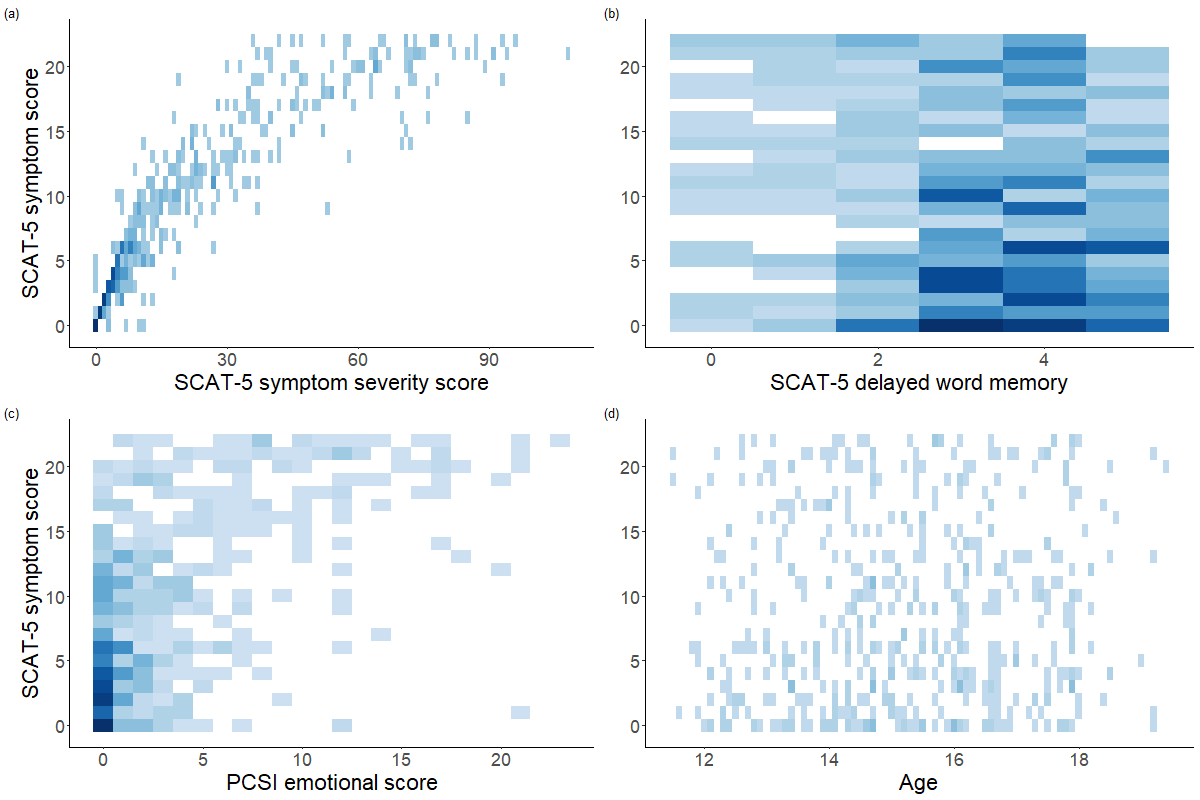}}
\caption{Density scatter plots of network output SCAT-5 symptom score vs. inputs (a) SCAT-5 symptom severity, (b) SCAT-5 delayed memory, (c) PCSI emotional score, and (d) age. Dark blue indicates higher density of observations.}
\label{fig:NN-SCAT}
\end{center}
\vskip -0.2in
\end{figure}

\begin{figure}[ht]
\vskip 0.2in
\begin{center}
\centerline{\includegraphics[width=\columnwidth]{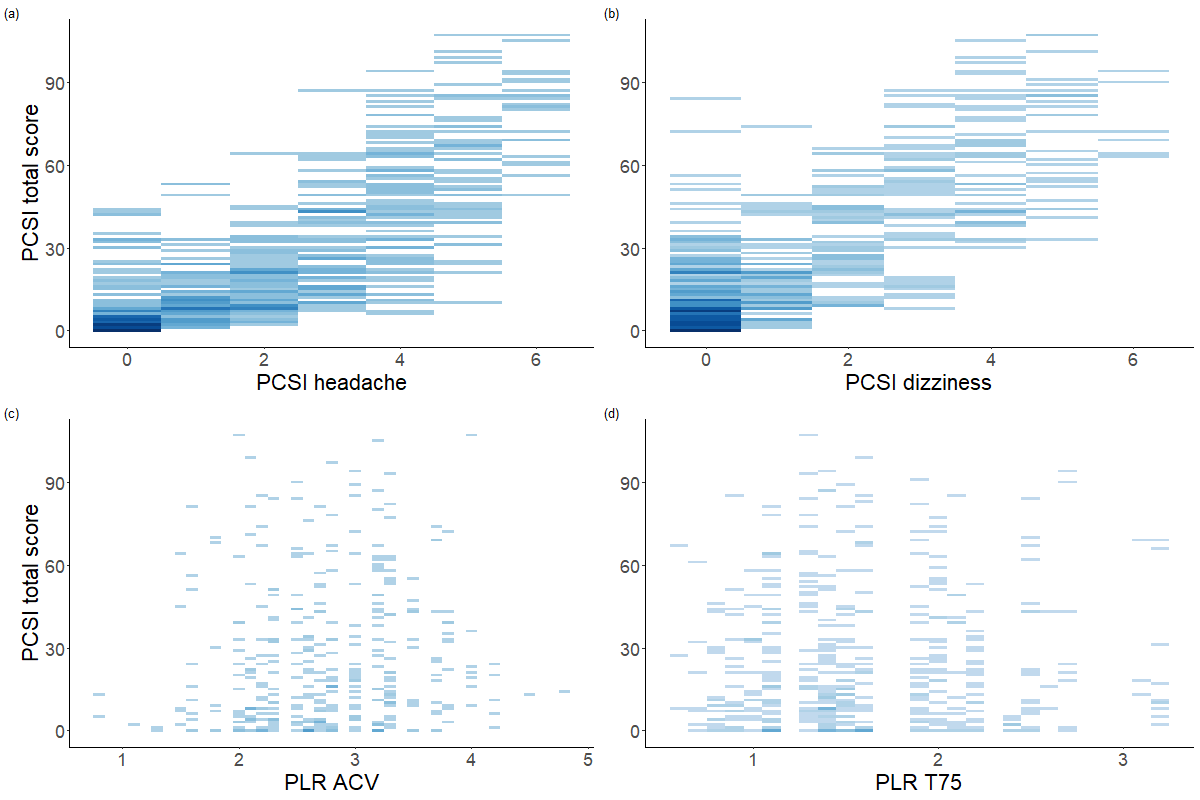}}
\caption{Density scatter plots of network output PCSI total score vs. inputs (a) PCSI headache, (b) PCSI dizziness, (c) PLR ACV, and (d) PLR T75. Dark blue indicates higher density of observations.}
\label{fig:NN-PCSI}
\end{center}
\vskip -0.2in
\end{figure}

Tests for nonlinearity and association were conducted for each input in each network. The p-values are reported in Table \ref{tab:CHOP_results}. In NN-SCAT, the tests suggest that SCAT-5 symptom severity is nonlinearly associated with SCAT-5 symptom score [nonlinearity: $p<0.001$; association: $p<0.001$]. Symptom score and symptom severity score are highly related measures as they are calculated from assessing the same 22 symptoms. However, the number of symptoms and the corresponding severity do not increase at the same rate. This nonlinear relationship is clearly visible in Figure \ref{fig:NN-SCAT}(a). Additionally, the tests suggest that PCSI emotional score is associated with symptom score [$p<0.001$], but there is not evidence that the association is nonlinear [$p=0.912$]. This result makes sense given that both metrics measure the presence of symptoms. The other two inputs, SCAT-5 delayed memory and age, do not show evidence of association with symptom score, an expected result given the even spread of values in Figure \ref{fig:NN-SCAT} (b) and (d). In NN-PCSI, the tests indicate that both headache and dizziness are strongly associated with total score [headache: $p<0.001$; dizziness: $p<0.001$] but neither association is nonlinear [headache: $p=0.642$, dizziness: $p=0.158$]. The tests confirm known linear associations; headache and dizziness are two of the elements summed to calculate total score. The linear associations are evident in Figure \ref{fig:NN-PCSI} (a) and (b). There is not evidence of an association between PCSI total score and either ACV [$p=0.100$] or T75 [$p=0.442$], an expected result given that PLR metrics measure a different dimension of concussion than symptoms and Figure \ref{fig:NN-PCSI} (c) and (d) shows no distinguishable relationships between the metrics.

\begin{table}[ht]
\caption{P-values for neural network permutation tests of nonlinearity and association for pediatric concussion measures. Tests are conducted for each of four features in two networks (NN-SCAT and NN-PCSI) at the 5\% level.}
\label{tab:CHOP_results}
\vskip 0.15in
\begin{center}
\begin{small}
\begin{sc}
\begin{tabular}{lrr}
\toprule
 & Nonlinearity & Association \\ 
\midrule
NN-SCAT \\
\hspace{1mm} Symptom severity & $<$0.001 & $<$0.001 \\
\hspace{1mm} Delayed memory & 0.298 & 0.702 \\
\hspace{1mm} Emotional score & 0.912 & $<$0.001 \\
\hspace{1mm} Age & 0.350 & 0.428 \\
NN-PCSI \\
\hspace{1mm} Headache & 0.642 & $<$0.001 \\
\hspace{1mm} Dizziness & 0.158 & $<$0.001 \\
\hspace{1mm} ACV & 0.442 & 0.100 \\
\hspace{1mm} T75 & 0.600 & 0.442 \\
\bottomrule
\end{tabular}
\end{sc}
\end{small}
\end{center}
\vskip -0.1in
\end{table}

\subsection{Testing genetic links to Parkinson's disease} 
The AMP PD database contains whole genome sequencing (WGS) and clinical data on subjects combined from four multicenter observational studies \citep{iwaki2021accelerating}. Several genes contain mutations known to be associated with the development of Parkinson's disease (PD), but the links between these genetic risk factors and PD are complex. We consider three genes with studied links to PD: SNCA \citep{polymeropoulos1997mutation}, SPPL3 \citep{zhang2022dysregulation}, and PLXNA4 \citep{schulte2013rare}. We include two genes linked to other conditions for comparison: HBB and CD4.

We analyze a set of 2890 subjects with WGS data; 1760 sucjects have a PD diagnosis. For each gene, we calculate the mean minor allele count across all SNPs as a summary measure. We train a one-layer network with 20 nodes and sigmoid activation using the five genetic predictors as well as age and sex to predict PD case status. We employ stochastic gradient descent with an initial learning rate of 0.06, which decays by 1.5\% at each epoch, and $L_2$ regularization with $\lambda=0.002$. The network trains for 125 epochs.

In addition to the neural network permutation test for association, we fit a linear model and a GAM to the data and conduct significance testing. The p-values are reported in Table \ref{tab:AMP}. All three methods suggest SNCA and PLXNA4 are significantly associated with PD, while none of the tests have sufficient evidence to suggest that SPPL3 is linked to PD. As expected, none of the methods find a significant association between PD and the HBB and CD4 genes.

\begin{table}[ht]
\caption{P-values for linear model (LM), GAM, and neural network (NN) tests for association for genetic predictors of PD. Tests are conducted for each feature at the 5\% level.}
\label{tab:AMP}
\vskip 0.15in
\begin{center}
\begin{small}
\begin{sc}
\begin{tabular}{lrrr}
\toprule
 & LM & GAM & NN \\ 
\midrule
SNCA & 0.001 & 0.003 & $<$0.001 \\
PLXNA4 & 0.014 & 0.013 & 0.010 \\
SPPL3 & 0.058 & 0.102 & 0.086\\
HBB & 0.652 & 0.292& 0.756 \\
CD4 & 0.919 & 0.873 & 0.410 \\
\bottomrule
\end{tabular}
\end{sc}
\end{small}
\end{center}
\vskip -0.1in
\end{table}

\section{Discussion}
\label{section:discussion}

In this article, we introduce a flexible permutation-based approach to hypothesis testing for neural networks. Our proposed tests utilize the partial derivative function of a network output with respect to specific inputs to evaluate associations between predictors and outcomes. There are several advantages to addressing the neural network explainability problem from the perspective of hypothesis testing. First, testing accounts for the overall network behavior rather than relying on local explanations at the individual prediction level. A global approach to network interpretation can better capture the nature of associations between predictors and outcomes. Second, feature importance methods provide relative rankings of predictors, but testing offers an objective interpretation of the significance of predictors analogous to inference conducted in classical statistical modeling like regression. Third, testing places no restrictions on network structure, allowing the data, rather than the need for explainability, to determine the optimal size and architecture. 

Basing our hypothesis testing framework on the partial derivatives allows for the method to be flexibly applied to general network structures. In this paper, we have implemented the tests in small, feed-forward networks which are appropriate to the scale and complexity of our data. However, the tests can be employed in any architecture where the partial derivatives can be calculated. For complex architectures where analytically deriving the partial derivatives proves difficult, it may be possible to approximate them in order to conduct testing. Despite the flexibility it affords, the permutation-based approach is computationally intensive, which limits the practicality of implementing the proposed tests for very large or complex networks. In these settings, an asymptotic test based on the partial derivatives would be an excellent alternative. However, while deriving an analytic form of the distribution of the partial derivative function is feasible, verifying it empirically is not a straightforward task. Many of the existing theoretical results for neural network parameters require that the network reach an optimal global solution, a condition that in practice, may be difficult to achieve or impossible to verify. Additionally, since the optimal network fit is achieved through numerical optimization rather than a closed-form solution, empirical estimates of the variance of the partial derivative must account for added variability due to training, especially when using methods such as stochastic gradient descent. Lastly, deriving an analytic form of the variance of the partial derivative requires the use of linear approximations to nonlinear functions. It is difficult to quantify the extent to which these approximations may bias the estimate or to study the conditions necessary to ensure a reasonable estimate. Together, these considerations make the development and implementation of an asymptotic test a rather challenging task. Empirical tests are therefore a useful alternative, especially for moderately-sized networks.

\bibliography{Mandel-permtest}
\bibliographystyle{icml2023}

\newpage
\appendix
\onecolumn
\section{Appendix}
\label{section:appendix}
The following outlines the testing procedures in detail.

\begin{algorithm}
\caption{Nonlinearity Test}
\begin{algorithmic}[1]
	\STATE Train a neural network on observed data $(\*X, \*y)$.
	\STATE Calculate the partial derivatives $\frac{\partial \mu_i}{\partial x_{ij}}$ at every observed value of $\*X_i$.
	\STATE Calculate the residuals of the partial derivatives from their mean:	$r_{ij} = \frac{\partial \mu_i}{\partial x_{ij}} -  \hat{c}$, where $\hat{c} = \frac{1}{n} \sum_{i=1}^n \frac{\partial \mu_i}{\partial x_{ij}}$.
	\STATE Fit a $q$-dimensional smooth function $s(t) = \*b(t)^T \*\theta$ to the $n$ residuals, where $\*b(t)$ are the basis functions and $\*\theta$ are the corresponding coefficients.
	\STATE Compute the test statistic $T=\frac{1}{q}\sum_{l=1}^q \hat{\theta}_l^2$.
	\STATE Fit a GAM to $(\*X, \*y)$, restricting $\*X_j$ to a linear term, and calculate the model residuals, $\*R$. Let $\hat{f}(\cdot)$ denote the estimated GAM.
	\STATE Permute the residuals and calculate $\*y^{(b)} = \hat{f}(\*X) + \*R^{(b)}$.
	\STATE Train a network on the permuted data $(\*X, \*y^{(b)})$, calculate the partial derivatives at every observed value of $\*x_i$, and calculate the residuals of the partial derivatives as in step 3.
	\STATE Fit a $q$-dimensional smooth function $s^{(b)}(t) = \*b(t)^T \*\theta^{(b)}$ to the $n$ residuals of the partial derivatives. Compute $T^{(b)}=\frac{1}{q}\sum_{l=1}^q \left(\hat{\theta}_l^{(b)} \right)^2$.
	\STATE Repeat steps 7-9 to obtain $T^{(1)},...,T^{(b)}$.
	\STATE Compute the p-value $p=\frac{1}{B} \sum_{b=1}^B I(T^{(b)} \geq T)$.
\end{algorithmic}
\end{algorithm}

\begin{algorithm}
\caption{Association Test}
\begin{algorithmic}[1]
	\STATE Train a neural network on observed data $(\*X, \*y)$.
	\STATE Calculate the partial derivatives $\frac{\partial \mu_i}{\partial x_{ij}}$ at every observed value of $\*x_i$.
	\STATE Compute the test statistic $T=\frac{1}{n}\sum_{i=1}^n \left( \frac{\partial \mu_i}{\partial x_{ij}} \right)^2$.
	\STATE Permute the observed values of $\*X_j$ such that all columns of the new predictor matrix $\*X^{(b)}$ are identical to $\*X$ except the $j$th.
	\STATE Train the network on $(\*X^{(b)}, \*y)$, calculate $\frac{\partial \mu_i}{\partial x^{(b)}_{ij}}$ for $i=1,...,n$, and compute $T^{(b)} = \frac{1}{n}\sum_{i=1}^n \left( \frac{\partial \mu_i}{\partial x_{ij}^{(b)}} \right)^2$.
	\STATE Repeat steps 4-5 to obtain $T^{(1)},...,T^{(b)}$.
	\STATE Compute the p-value $p=\frac{1}{B} \sum_{b=1}^B I(T^{(b)} \geq T)$.
\end{algorithmic}
\end{algorithm}

\end{document}